\documentclass[APA,STIX1COL]{WileyNJD-v2}
\usepackage{subcaption}

\articletype{Article Type}%

\received{ }
\revised{ }
\accepted{ }

\newcommand{\modelname}{RENeGe}
\newcommand{\modelnamesn}{RENeGe-sk}

\usepackage{tikz}
\usetikzlibrary{calc}

\tikzset{
	regnode/.style={
		circle,
		draw=black,
		fill=white,
		inner sep=2pt,   
		font=\small       
	}
}

\tikzset{
	edgelabel/.style={
		fill=blue!15,
		inner sep=0.7pt,      
		font=\tiny,  
		text=black
	}
}

\tikzset{
	lineedge/.style={
		draw=blue!60,
		line width=1pt,
		bend left=-50
	}
}

\tikzset{
	edgelabel1/.style={
		fill=blue!15,
		inner sep=2pt,      
		font=\scriptsize,  
		text=black
	}
}

\begin{document}

\title{Directional Asymmetry in Edge-Based Spatial Models via a Skew-Normal Prior}

\author[1]{Danna L. Cruz-Reyes}
\author[2]{Renato M. Assun\c{c}\~{a}o}
\author[3]{Reinaldo B. Arellano-Valle}
\author[4]{Rosangela H. Loschi}


\address[1]{\orgdiv{Departamento de Estadística}, 
\orgname{Universidad Nacional de Colombia}, 
\orgaddress{\city{Bogot\'{a}}, \country{Colombia}}}

\address[2]{\orgname{Esri}, 
\orgaddress{\city{Redlands}, \state{CA}, \country{USA}}; 
\orgdiv{Departamento de Ci\^{e}ncia da Computa\c{c}\~{a}o}, 
\orgname{Universidade Federal de Minas Gerais}, 
\orgaddress{\city{Belo Horizonte}, \state{MG}, \country{Brazil}}}

\address[3]{\orgdiv{Departamento de Estad\'{\i}stica}, 
\orgname{Pontif\'{i}cia Universidad Cat\'{o}lica}, 
\orgaddress{\city{Santiago},  \country{Chile}}}

\address[4]{\orgdiv{Departamento de Estat\'{\i}stica}, 
\orgname{Universidade Federal de Minas Gerais}, 
\orgaddress{\city{Belo Horizonte}, \state{MG}, \country{Brazil}}}

\corres{
\textsuperscript{*}Corresponding author: 
Danna L. Cruz-Reyes. 
\email{dlcruzr@unal.edu.co}
}


\abstract[Summary]{We introduce a skewed edge-based spatial prior,   named \modelnamesn{}, that extends the Gaussian RENeGe framework by incorporating directional asymmetry through a skew–normal distribution. Skewness is defined on the edge graph and propagated to the node space, aligning asymmetric behavior with transitions across neighboring regions rather than with marginal node effects. The model is formulated within the skew–normal framework and employs identifiable hierarchical priors together with low-rank parameterizations to ensure scalability. The skew–normal's stochastic representation is considered to facilitate the computational implementation.
Simulation studies show that \modelnamesn{} recovers compact, edge-aligned directional structure more accurately than symmetric Gaussian priors, while remaining competitive under irregular spatial patterns. An application to cancer incidence data in Southern Brazil illustrates how the proposed approach yields stable area-level estimates while preserving localized, directionally driven spatial variation.

}


\maketitle



\section{Introduction}

Spatial hierarchical models for areal data commonly represent residual geographic structure through latent random effects defined on the nodes of a neighborhood graph. In disease mapping and related small-area problems, this latent field captures spatial dependence beyond observed covariates and stabilizes estimates in regions with sparse counts. A standard choice is the conditional autoregressive (CAR) prior \citep{besag1974spatial}, which induces a joint Gaussian distribution for $\boldsymbol{\theta}=(\theta_1,\ldots,\theta_n)^\top$ through local conditional specifications. Under mild regularity conditions, a proper CAR model can be written as
\begin{equation}\label{eq:CAR_intro}
\boldsymbol{\theta}\sim \mathcal{N}_n\!\left(\boldsymbol{0},\,\tau_\theta^2(\boldsymbol{M}-\varsigma \boldsymbol{A})^{-1}\right),
\end{equation}
where $\boldsymbol{A}$ is the $n\times n$ adjacency matrix in areas, $\boldsymbol{M}=\mathrm{diag}(m_1,\dots,m_n)$ is the degree matrix, $\tau_\theta^2>0$ is a scale parameter, and $(\boldsymbol{M}-\varsigma \boldsymbol{A})$ is the precision matrix. The scalar $\varsigma$ is referred to as the spatial dependence parameter. Positive-definiteness holds for $\varsigma$ in the open interval determined by the extremal eigenvalues of $\boldsymbol{M}^{-1/2}\boldsymbol{A}\boldsymbol{M}^{-1/2}$. An important limiting case is the intrinsic CAR (ICAR) prior \citep{BYM}, obtained informally as $\varsigma\to 1$ (or through singular precision), which is widely used as an improper smoothing prior.

Despite their popularity and computational convenience, CAR/ICAR priors have well-documented shortcomings. \citet{wall_close_2004} highlighted several counterintuitive features: correlations between neighboring areas can vary systematically with node degree; regions with the same number of neighbors can have different marginal variances; and the resulting dependence can be sensitive to graph properties that are not related to substantive spatial mechanisms. These phenomena were later clarified by \citet{Assuncao}, who showed that the CAR-induced dependence reflects the \emph{global} spectral structure of the neighborhood graph (including higher-order paths), rather than only the first-order adjacency. Such issues are particularly problematic on heterogeneous graphs, motivating the exploration of alternative ways to encode neighborhood information in spatial priors.

\begin{figure*}[htb]
\centering
\resizebox{\textwidth}{!}{%
\begin{minipage}[t]{0.2\textwidth}
\centering
\begin{tikzpicture}
\node[anchor=south west, inner sep=0] at (-0.2,0.1)
{\includegraphics[width=3cm]{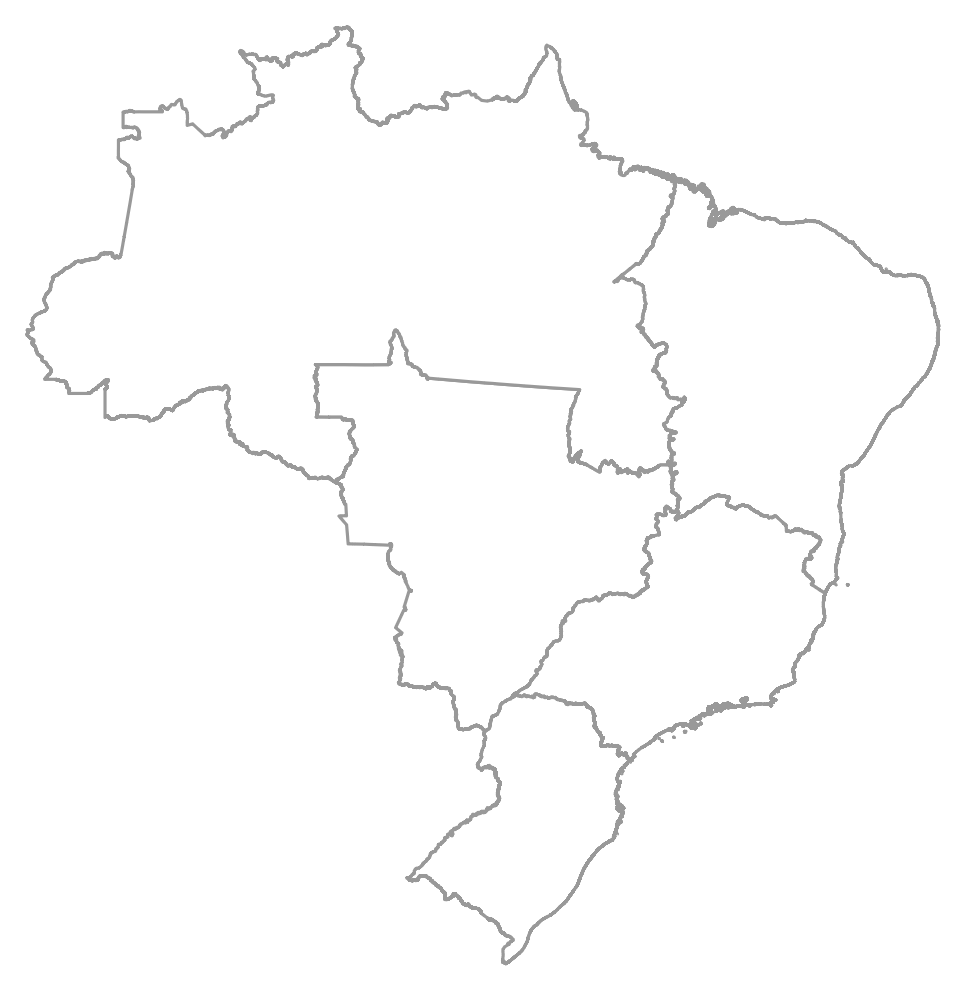}};

\node[regnode] (1) at (0.7,2.6) {1};
\node[regnode] (2) at (2.25,2.1) {2};
\node[regnode] (3) at (1.3,1.5) {3};
\node[regnode] (4) at (2,1.2) {4};
\node[regnode] (5) at (1.4,0.5) {5};

\draw (1)--(2);
\draw (1)--(3);
\draw (2)--(3);
\draw (2)--(4);
\draw (3)--(4);
\draw (4)--(5);
\draw (3)--(5);

\end{tikzpicture}

{\small (a) $\mathcal{G}$}
\end{minipage}
\hfill

\begin{minipage}[t]{0.2\textwidth}
\centering
\begin{tikzpicture}

\node[anchor=south west, inner sep=0, opacity=0.25] at (-0.2,-0.1)
{\includegraphics[width=3cm]{mapa}};

\node[regnode] (1) at (0.7,2.4) {1};
\node[regnode] (2) at (2.25,1.8) {2};
\node[regnode] (3) at (1.3,1.2) {3};
\node[regnode] (4) at (2,1) {4};
\node[regnode] (5) at (1.4,0.3) {5};

\draw (1)--node[edgelabel,pos=0.55]{12}(2);
\draw (1)--node[edgelabel,pos=0.55]{13}(3);
\draw (2)--node[edgelabel,pos=0.55]{23}(3);
\draw (2)--node[edgelabel,pos=0.55]{24}(4);
\draw (3)--node[edgelabel,pos=0.55]{34}(4);
\draw (4)--node[edgelabel,pos=0.55]{45}(5);
\draw (3)--node[edgelabel,pos=0.55]{35}(5);

\end{tikzpicture}

{\small (b) $\mathcal{E}$}
\end{minipage}
\hfill

\begin{minipage}[t]{0.2\textwidth}
\centering
\begin{tikzpicture}

\node[anchor=south west, inner sep=0, opacity=0.25] at (-0.2,-0.1)
{\includegraphics[width=3cm]{mapa}};

\node[opacity=0] (1) at (1.3,2.4) {1};
\node[opacity=0] (2) at (2.25,1.8) {2};
\node[opacity=0] (3) at (1.3,1.2) {3};
\node[opacity=0] (4) at (2,1) {4};
\node[opacity=0] (5) at (1.4,0.3) {5};

\node[edgelabel] (A12) at ($(1)!0.5!(2)$) {12};
\node[edgelabel] (A13) at ($(1)!0.5!(3)$) {13};
\node[edgelabel] (A23) at ($(2)!0.5!(3)$) {23};
\node[edgelabel] (A24) at ($(2)!0.5!(4)$) {24};
\node[edgelabel] (A34) at ($(3)!0.5!(4)$) {34};
\node[edgelabel] (A45) at ($(4)!0.5!(5)$) {45};
\node[edgelabel] (A35) at ($(3)!0.5!(5)$) {35};

\draw[blue!60, line width=0.8pt, bend left=-50] (A12) to (A13);
\draw[blue!60, line width=0.8pt, bend right=-30] (A12) to (A23);
\draw[blue!60, line width=0.8pt, bend left=-20] (A13) to (A23);
\draw[blue!60, line width=0.8pt, bend left=40] (A23) to (A24);
\draw[blue!60, line width=0.8pt, bend right=30] (A23) to (A34);
\draw[blue!60, line width=0.8pt, bend left=30] (A24) to (A34);
\draw[blue!60, line width=0.8pt, bend left=25] (A34) to (A45);
\draw[blue!60, line width=0.8pt, bend right=20] (A34) to (A35);
\draw[blue!60, line width=0.8pt, bend left=-45] (A35) to (A45);

\end{tikzpicture}

{\small (c) $\mathcal{L(G)}$}
\end{minipage}
}
\caption{
Schematic representation of an areal map (left), its induced neighborhood graph (center), and the corresponding line graph (right).
Nodes in the line graph represent adjacencies in the original map, enabling spatial dependence and directional effects to be modeled directly on edges.
}
\label{fig:graph_representation}

\end{figure*}

Figure~\ref{fig:graph_representation} provides a schematic illustration of the representation underlying edge-based spatial models. An areal map induces a neighborhood graph whose nodes correspond to regions and whose edges represent shared boundaries. The associated line graph $\mathcal{L}(\mathcal{G})$ is obtained by treating each edge of the original graph as a node, with adjacency defined by shared endpoints. In this representation, spatial transitions are encoded directly on adjacencies rather than on regions themselves, making the edge graph a natural domain for modeling boundary-driven and directional effects. The definitions and properties of this construction are given in Section~2 of \citet{renege2024}.


This perspective underlies the \modelname{} model proposed by \citet{renege2024}, which constructs spatial dependence by assigning latent effects to the edges and then projecting them to the nodes. Let $\mathcal{G}=(\mathcal{V},\mathcal{E})$ be the area graph with $|\mathcal{V}|=n$ and $|\mathcal{E}|=p$. Let $\boldsymbol{C}\in\{0,1\}^{n\times p}$ be the node--edge incidence matrix, and $\boldsymbol{A}_e$ and $\boldsymbol{M}_e$, respectively, denote the adjacency and degree matrices on the line graph $\mathcal{L}(\mathcal{G})$ (the graph whose nodes are the edges of $\mathcal{G}$). A Gaussian \modelname{} baseline specifies an edge-level latent vector $\boldsymbol{\varepsilon}\in\mathbb{R}^p$  and assumes that
\begin{equation}\label{eq:\modelname{}_gauss_intro}
\boldsymbol{\rho}\sim \mathcal{N}_p\!\bigl(\boldsymbol{0},\,\boldsymbol{\Omega}\bigr),
\qquad \boldsymbol{\Omega}=\bigl(\boldsymbol{M}_e-\gamma \boldsymbol{A}_e\bigr)^{-1},
\qquad \boldsymbol{\theta}=\boldsymbol{C}\boldsymbol{\rho},
\end{equation}

with $\gamma$ restricted so that $\boldsymbol{\Omega}$ is positive definite. By making the line graph $\mathcal{L}(\mathcal{G})$ the primary object governing dependence, \modelname{} yields node-level covariances that respect shared-edge combinatorics and mitigate several CAR/ICAR anomalies.

 The Gaussian specification in \eqref{eq:CAR_intro} and \eqref{eq:\modelname{}_gauss_intro} imposes symmetry and thin tails. 
 However, in disease mapping and related applications, latent spatial structure may display asymmetric behavior and occasional extreme deviations; these features can be consequential when risk contrasts concentrate along a subset of edges.  Skewed models for spatial data  can be found, for instance, in  \cite{Zhang2010}, \cite{hosseini2011approximate}, \cite{Genton2012}, \cite{pratesdeylachos2012}, \cite{rantini2021fernandez}, and \cite{AYALEW2024}.  
 
 Spatial asymmetry is fundamentally a property of transitions across neighboring regions, rather than of regions in isolation. Encoding skewness directly on the node space therefore risks confounding directional effects with local variance heterogeneity or irregular neighborhood structure. By defining skewness on the edge graph and projecting it to nodes, the proposed model aligns asymmetry with the spatial mechanism through which it acts—differences across shared edges—while preserving the stability and interpretability of edge-based Gaussian smoothing. This motivates extending edge-based priors beyond the Gaussian family.

In this paper, we propose \modelnamesn{}, a skew-normal extension of \modelname{} that introduces \emph{directional asymmetry} directly in the edge space and then projects it to nodes.  The model is built within the skew-normal (SN) framework \citep{arellano2006, azzalini2005}, considering its stochastic representation, which allows efficient Bayesian computation and scalable low-rank parameterizations \citep{hughes2013dimension} of the skewness direction. Intuitively, the Gaussian \modelname{} component controls smooth propagation on $\mathcal{L}(\mathcal{G})$, while the skewness component adds an interpretable directional perturbation concentrated on selected edges.

Our contributions are threefold. First, we introduce the \modelnamesn{} prior by specifying a multivariate skew-normal law on the edge-based latent vector and deriving the induced marginal distribution on the node space. Second, we provide a fully Bayesian hierarchical specification with identifiable priors for both the edge-graph dependence parameters and the low-rank skewness coefficients, along with practical simulation and inference strategies. Third, through controlled experiments and an application to cancer mortality in Brazil, we show that \modelnamesn{} can recover compact, edge-aligned directional patterns that symmetric Gaussian priors tend to smooth away, while remaining competitive under irregular latent structure.

The remainder of the article is organized as follows. Section~\ref{sec:\modelname_sn} defines the \modelnamesn{} prior and derives its key properties. Section~\ref{sec:sim_\modelname_sn} presents a simulation using the auxiliary-variable representation. Section~\ref{sec:bayes_\modelname_sn} introduces the full hierarchical model within a Poisson log-linear framework. Section~\ref{sec:requirements_skew} discusses when an edge-based prior can be regarded as genuinely skew-normal and illustrates the resulting directional behavior in a synthetic example. Section~\ref{aplicacion} presents the empirical application, and Section~\ref{sec:conclusions} concludes with future directions.


\section{ An Edge-Based Skew-Normal Prior for Spatial Fields}
\label{sec:\modelname_sn}

The goal of this section is to define a spatial prior that combines the edge-based dependence structure of the Gaussian \modelname{} model with a principled mechanism for directional asymmetry.  We achieve this by assuming a multivariate skew--normal prior distribution \citep{arellano2006} directly on the latent edge effects $\boldsymbol{\varepsilon}$ and then projecting this structure to the node space through the incidence matrix. This construction preserves the geometric interpretation of edge-based smoothing while allowing departures from symmetry that are aligned with specific subsets of edges.

We start by explaining the edge-based skew--normal specification. 
Let $\boldsymbol{\rho}\in\mathbb{R}^p$ denote the latent vector defined on the edges of the spatial graph, where $p=|\mathcal{E}|$. In the Gaussian \modelname{} model in (\ref{eq:\modelname{}_gauss_intro}), $\boldsymbol{\rho}$ follows a zero-mean multivariate normal distribution with precision induced by the line graph $\mathcal{L}(\mathcal{G})$. To introduce directional asymmetry, we replace this Gaussian law with a multivariate skew--normal distribution whose skewness direction is itself defined on the edge space. Specifically, we assign
\begin{equation} \label{EqReSN}
\boldsymbol{\rho}\;\sim\; \operatorname{SN}_p\!\left(
-\,b\,\boldsymbol{\eta},\;
\sigma_\theta^2\,(M_e-\gamma A_e)^{-1};\;
\boldsymbol{\eta}
\right),
\qquad b=\sqrt{\tfrac{2}{\pi}},    
\end{equation}
where $M_e$ and $A_e$ are, respectively, the degree and adjacency matrices on the line graph, $\gamma$ controls the strength of edge-to-edge dependence, and $\boldsymbol{\eta}\in\mathbb{R}^p$ is a vector encoding the direction and magnitude of skewness on edges. The centering term $-b\,\boldsymbol{\eta}$ ensures that the prior mean of $\boldsymbol{\rho}$ is zero, so that asymmetry affects higher moments but not the first moment of the spatial field.

 The skew--normal prior in (\ref{EqReSN}) admits a convenient stochastic representation that plays a central role in both interpretation and computation. Such a prior can be hierarchically  represented as 
\begin{equation} \label{EqSR}
\begin{cases}
\boldsymbol{\rho} \;=\; -\,b\,\boldsymbol{\eta} \;+\; \boldsymbol{\eta}\,U \;+\; \boldsymbol{\varepsilon},\\[2mm]
U \sim \mathrm{TN}_+(0,1)\quad (\text{equivalently } U=|Z|,\ Z\sim\mathcal N(0,1)),\\[1mm]
\boldsymbol{\varepsilon} \sim \mathcal N_p\!\bigl(\mathbf 0,\ \sigma_\theta^2\,(M_e-\gamma A_e)^{-1}\bigr),
\end{cases}
\end{equation}
where $U$ is a latent half--normal variable independent of the Gaussian component $\boldsymbol{\varepsilon}$. Conditional on $U$, the edge effects $\boldsymbol{\rho}$ are Gaussian with mean $-b\,\boldsymbol{\eta}+\boldsymbol{\eta}U$ and covariance governed by the same edge-based precision as in the Gaussian \modelname{} model. This stochastic representation clarifies the role of skewness: the Gaussian term $\boldsymbol{\varepsilon}$ controls smooth propagation across adjacent edges, while the scalar $U$ modulates a global directional shift along $\boldsymbol{\eta}$. When $\boldsymbol{\eta}=\mathbf 0$, the skew--normal prior collapses exactly to the Gaussian \modelname{} specification.
%

We now describe how edge-level effects induce a spatial field on the nodes.  The latent spatial field in the areas is obtained by projecting edge effects onto the nodes through  the incidence matrix $C\in\mathbb{R}^{n\times p}$,
\[
\boldsymbol{\theta}=C\,\boldsymbol{\rho}.
\]
Substituting the auxiliary-variable representation yields
\[
\boldsymbol{\theta}
= -\,b\,C\boldsymbol{\eta}
+ C\boldsymbol{\eta}\,U
+ C\boldsymbol{\varepsilon}.
\]
Since $C\boldsymbol{\varepsilon}$ is Gaussian with covariance
\[
\sigma_\theta^2\,C\,(M_e-\gamma A_e)^{-1}C^\top,
\]
the conditional distribution of $\boldsymbol{\theta}$ given $U$ is multivariate normal,
\[
\boldsymbol{\theta}\mid U \sim
\mathcal N_n\!\Big(
-\,b\,C\boldsymbol{\eta}+C\boldsymbol{\eta}\,U,\;
\sigma_\theta^2\,C\,(M_e-\gamma A_e)^{-1}C^\top
\Big),
\qquad
U\sim\mathrm{TN}_+(0,1).
\]

Marginally, the node-level spatial field follows a multivariate skew--normal distribution,
\[
\boldsymbol{\theta}\sim
\operatorname{SN}_n\!\Big(
-\,b\,C\boldsymbol{\eta}\ ;\
\sigma_\theta^2\,C\,(M_e-\gamma A_e)^{-1}C^\top\ ;\
C\boldsymbol{\eta}
\Big).
\]

Finally, the moments of the induced spatial field highlight how asymmetry augments the Gaussian edge-based structure. By construction, the prior mean of the spatial field is zero, 
\[
\mathbb{E}(\boldsymbol{\theta})=\mathbf 0,
\]
while the covariance decomposes as
\[
\mathbb{V}(\boldsymbol{\theta})
=
\left(1-\tfrac{2}{\pi}\right)
C\boldsymbol{\eta}\boldsymbol{\eta}^\top C^\top
\;+\;
\sigma_\theta^2\,C\,(M_e-\gamma A_e)^{-1}C^\top.
\]
The second term coincides with the Gaussian \modelname{} covariance and encodes symmetric spatial dependence through the line graph. The first term is a rank--one positive semi-definite component that introduces directional variability aligned with the projected skewness vector $C\boldsymbol{\eta}$. This decomposition makes explicit how \modelnamesn{} augments the Gaussian edge-based model with a parsimonious and interpretable form of asymmetry.

In summary, \modelnamesn{} preserves the interpretation of the neighborhood and computational advantages of \modelname{} prior while extending it to represent directional spatial behavior that symmetric Gaussian models cannot capture.

\section{Simulation from the Edge-Based Skew-Normal Prior}
\label{sec:sim_\modelname_sn}

 The simulation of the \modelnamesn{} prior is carried out conveniently using the stochastic representation of the skew--normal distribution shown in Section~\ref{sec:\modelname_sn}. The representation in expression (\ref{EqSR}) expresses the latent spatial field as a Gaussian component perturbed by a single latent half-normal variable, which greatly simplifies both prior simulation and posterior computation.

 Let $b=\sqrt{2/\pi}$. Denote by $\boldsymbol{\eta}\in\mathbb{R}^{p}$  the skewness direction on the edge space, and let $C\in\mathbb{R}^{n\times p}$ be the node--edge incidence matrix. Recall that the edge--based latent vector $\boldsymbol{\rho}$ admits the stochastic representation
\[
\boldsymbol{\rho}
= -\,b\,\boldsymbol{\eta}
+ \boldsymbol{\eta}\,U
+ \boldsymbol{\varepsilon},
\qquad
U=|Z|,\ Z\sim\mathcal N(0,1),
\qquad
\boldsymbol{\varepsilon}\sim\mathcal N_p\!\bigl(\mathbf 0,\ \sigma_\theta^2(M_e-\gamma A_e)^{-1}\bigr).
\]

\noindent Projecting to the node space yields
\[
\boldsymbol{\theta}
= C\boldsymbol{\rho}
= -\,b\,C\boldsymbol{\eta}
+ C\boldsymbol{\eta}\,U
+ C\boldsymbol{\varepsilon}.
\]
Conditionally on $U$, the spatial field $\boldsymbol{\theta}$ is multivariate Gaussian with mean $\boldsymbol{\mu}_\theta = -\,b\,C\boldsymbol{\eta}$
and covariance matrix $\boldsymbol{\Sigma}_\theta
= \sigma_\theta^2\,C\,(M_e-\gamma A_e)^{-1}C^\top$.
The latent variable $U$ introduces a global directional perturbation aligned with $C\boldsymbol{\eta}$, while $\boldsymbol{\Sigma}_\theta$ coincides with the Gaussian \modelname{} covariance on the node space.

\subsection*{Simulation algorithm}

Sampling from the \modelnamesn{} prior for $\boldsymbol{\theta}$ proceeds as follows:

\begin{enumerate}
\item Compute the mean vector $\boldsymbol{\mu}_\theta = -\,b\,C\boldsymbol{\eta}$ and the covariance matrix $\boldsymbol{\Sigma}_\theta = \sigma_\theta^2\,C\,(M_e-\gamma A_e)^{-1}C^\top$.

\item Draw a scalar $Z\sim\mathcal N(0,1)$ and set $U=|Z|$.

\item Draw $\boldsymbol{\varepsilon}_\theta \sim \mathcal N_n(\mathbf 0,\boldsymbol{\Sigma}_\theta)$.

\item Set $\boldsymbol{\theta}
= \boldsymbol{\mu}_\theta
+ C\boldsymbol{\eta}\,U
+ \boldsymbol{\varepsilon}_\theta.$
\end{enumerate}

This procedure requires only standard Gaussian sampling and a single factorization of the covariance matrix $\boldsymbol{\Sigma}_\theta$. It avoids direct sampling from truncated multivariate distributions and is numerically stable even for moderately large graphs.

When $\boldsymbol{\eta}=\mathbf 0$, the algorithm reduces exactly to simulation from the Gaussian \modelname{} prior. More generally, the magnitude and orientation of $C\boldsymbol{\eta}$ determine the strength and direction of asymmetry in the simulated field. This makes the simulation scheme especially useful for generating controlled synthetic examples in which directional spatial behavior is prescribed on a subset of edges.

\section{The \modelnamesn~model for count data}
\label{sec:bayes_\modelname_sn}

We now embed the \modelnamesn{} prior within a full Bayesian hierarchical model for areal count data. The formulation extends the Gaussian \modelname{} specification by introducing a skew--normal perturbation on the edge graph, which induces directional asymmetry in the node--level spatial field. The resulting model retains the interpretability of edge--based smoothing while allowing departures from symmetry to be learned from the data.

Let $Y_i$ denote the observed count in area $i$, for $i=1,\ldots,n$, and let $E_i$ be a known offset or expected count. Conditional on a linear predictor $\eta_i$, we assume a Poisson log--linear model,
\[
Y_i \mid \eta_i \sim \text{Poisson}\!\left(\exp(\eta_i)\right),
\qquad
\eta_i = \alpha + X_i^\top \beta + \log(E_i) + \theta_i,
\]
where $\alpha$ is an intercept, $X_i$ is a vector of covariates with regression coefficients $\beta$, and $\theta_i$ represents the latent spatial effect in area $i$.

Spatial dependence between the areas is indirectly introduced through a latent vector $\boldsymbol{\rho}\in\mathbb{R}^p$ defined on the edges of the neighborhood graph. Following Section~\ref{sec:\modelname_sn}, we assign a skew--normal prior on the edge space,
\[
\boldsymbol{\rho}
\;\sim\;
\operatorname{SN}_p\!\left(
-\,\sqrt{\tfrac{2}{\pi}}\,\boldsymbol{\eta},\;
\sigma_\theta^2\,(M_e-\gamma A_e)^{-1};\;
\boldsymbol{\eta}
\right),
\]
where $M_e$ and $A_e$ are, respectively, the degree and adjacency matrices of the line graph $\mathcal{L}(\mathcal{G})$,  $\gamma$ controls the strength of edge--to--edge dependence, $\sigma_\theta^2$ is a global scale parameter, and $\boldsymbol{\eta}\in\mathbb{R}^p$ encodes the direction and magnitude of skewness on edges.


The spatial effect at the areal level is obtained by projecting edge effects to nodes using the incidence matrix $C\in\mathbb{R}^{n\times p}$,
\[
\boldsymbol{\theta} = C\,\boldsymbol{\rho}.
\]
Considering the stochastic representation of $\boldsymbol{\rho}$   given in (\ref{EqSR}) it follows that
\[
\boldsymbol{\theta}
=
-\,\sqrt{\tfrac{2}{\pi}}\,C\boldsymbol{\eta}
+ C\boldsymbol{\eta}\,U
+ C\boldsymbol{\varepsilon}.
\]
Conditionally on $U$, the spatial field is Gaussian,
\[
\boldsymbol{\theta}\mid U
\sim
\mathcal N_n\!\Big(
-\,\sqrt{\tfrac{2}{\pi}}\,C\boldsymbol{\eta}
+ C\boldsymbol{\eta}\,U,\;
\sigma_\theta^2\,C(M_e-\gamma A_e)^{-1}C^\top
\Big),
\]
which makes explicit how the asymmetry in the edge graph propagates to the node space through the linear map $C$.

We complete the hierarchical specification by assigning few informative prior distributions to regression parameters,
\[
\alpha \sim \mathcal N(0,10),
\qquad
\beta \sim \mathcal N(\mathbf 0,5^2 I),
\]
and hyperpriors governing spatial dependence,
\[
\gamma \sim \mathcal U(0,1),
\qquad
\sigma_\theta^{-2} \sim \mathrm{Gamma}(a_\tau,b_\tau).
\]

To ensure identifiability and parsimony, we impose a hierarchical prior on the skewness vector,
\[
\boldsymbol{\eta} = \sigma_\eta\,\boldsymbol{\eta}_{\text{raw}},
\qquad
\boldsymbol{\eta}_{\text{raw}} \sim \mathcal N_p(\mathbf 0, I),
\qquad
\sigma_\eta \sim \mathrm{Half\mbox{-}Normal}(0,1).
\]
This formulation induces an isotropic prior over skewness directions on the edge graph, while the scalar parameter $\sigma_\eta$ controls the overall strength of asymmetry. When $\sigma_\eta$ is shrunk toward zero, the model reduces smoothly to the Gaussian \modelname{} specification.

  The stochastic representation of the Skew-normal distribution allows for an efficient posterior computation and facilitates joint inference on spatial smoothing, directional asymmetry, and regression effects. The proposed model is implemented in Stan and the code are available in GitHub repository at \url{https://github.com/DannaCruz/RENeGe-sk}.


\section{When Is a Spatial Prior Genuinely Skew-Normal?}
\label{sec:requirements_skew}

Spatial asymmetry is inherently a property of transitions between neighboring regions, not of regions in isolation. Abrupt changes, directional gradients, and boundary-aligned effects arise when crossing specific adjacencies, rather than from marginal behavior at individual nodes. Encoding skewness directly on the node space therefore risks conflating directional effects with local variance heterogeneity or irregular neighborhood structure. By contrast, defining skewness on the edge graph aligns the source of asymmetry with the spatial mechanism through which it operates: differences across shared borders. In the \modelnamesn{} construction, the Gaussian component governs symmetric propagation of dependence across adjacent edges, while a low-rank skewness perturbation assigns directional weight to a subset of edges. Projection to the node space then yields an asymmetric field whose direction, location, and strength are jointly identifiable and interpretable. This separation preserves the interpretability and stability of Gaussian edge-based smoothing while allowing asymmetric behavior to emerge only where the graph structure supports it.

The presence of asymmetric marginal distributions in a latent spatial field is not, by itself, sufficient to characterize a spatial prior as skew--normal. In a genuine skew--normal construction, asymmetry arises from a coherent joint mechanism rather than from isolated irregularities at individual nodes. In particular, the joint distribution must admit a representation in which skewness is induced by a single truncated Gaussian component whose influence propagates systematically across the domain. This requirement is characteristic of the skew--normal(SN) family and ensures that skewness has a well-defined direction and magnitude at the multivariate level.

In the \modelnamesn{} model, this requirement naturally leads to the edge graph. Directional asymmetry in spatial fields is typically associated with abrupt changes that occur \emph{across adjacencies}, rather than within areas themselves. Encoding skewness directly on the node space risks confusing asymmetry with heterogeneous local variance or irregular neighborhood structure. By contrast, defining the skewness vector $\boldsymbol{\eta}\in\mathbb{R}^p$ on the edges of the graph aligns the source of asymmetry with the spatial mechanism through which it acts: transitions between neighboring regions.

Under the \modelnamesn{} construction, the edge-level latent vector $\boldsymbol{\rho}$ follows a skew--normal distribution whose Gaussian component is governed by the \modelname{} precision $(M_e-\gamma A_e)^{-1}$ on the line graph, while asymmetry is introduced through a rank--one perturbation in the direction $\boldsymbol{\eta}$. Projection to the node space via $\boldsymbol{\theta}=C\boldsymbol{\rho}$ yields a spatial field whose covariance decomposes into a symmetric component induced by edge-based smoothing and an additional low-rank term that captures directional variability. As shown in Section~\ref{sec:\modelname_sn}, this decomposition preserves the neighborhood interpretation of the Gaussian base model while augmenting it with interpretable asymmetry.

To illustrate the type of behavior generated by this mechanism, we consider a controlled synthetic example based on the microregions of southern Brazil. The underlying graph consists of $n=159$ nodes and their corresponding edges. We construct a latent field exhibiting a pronounced north--south gradient, where values increase gradually in the central region but rise sharply when crossing a specific band of edges separating the southern block from the remainder of the domain. This configuration is designed so that the dominant asymmetry is localized along a subset of adjacencies rather than distributed uniformly across nodes.

\begin{figure}[t]
\centering
\includegraphics[width=0.45\textwidth]{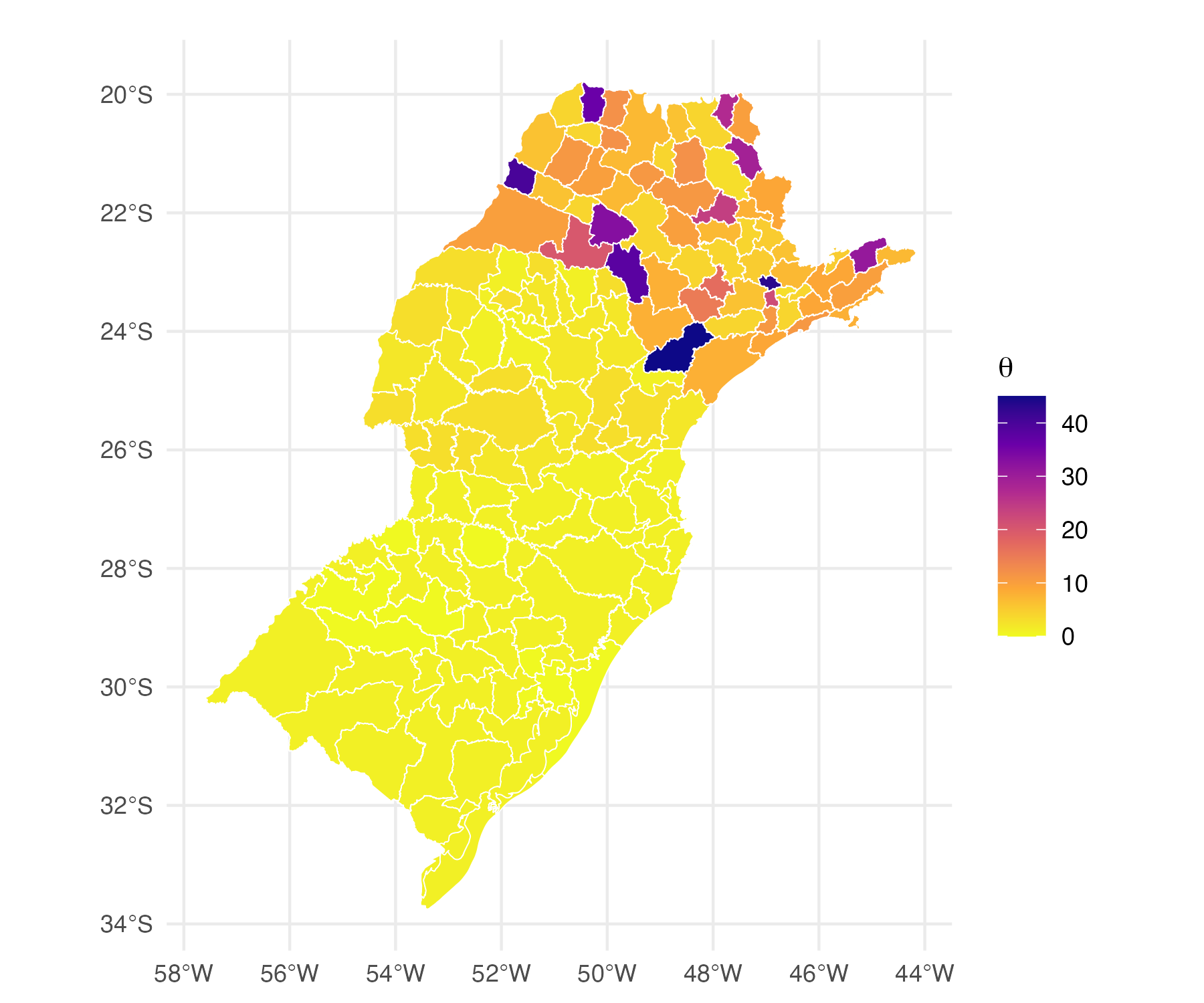}
\caption{Simulated latent field {$\theta$ exhibiting a }directional north--south increase generated through an edge-based Skew--Normal perturbation.}
\label{fig:Y_simul_tp}
\end{figure}

A purely Gaussian spatial prior, whether CAR or \modelname{}, can reproduce the overall smoothness of this surface but necessarily regularizes transitions symmetrically. As a result, abrupt changes tend to be diffused over a wide neighborhood. In contrast, by placing the skewness vector $\boldsymbol{\eta}$ on the edges corresponding to the central--southern interface, the \modelnamesn{} prior produces an asymmetric rise that is markedly sharper in one direction than in the other. The resulting latent field, displayed in Figure~\ref{fig:Y_simul_tp}, exhibits a directional transition that is aligned with the specified edge set.

\begin{figure}[t!]
\centering
\begin{subfigure}{0.45\textwidth}
\centering
\includegraphics[width=\linewidth]{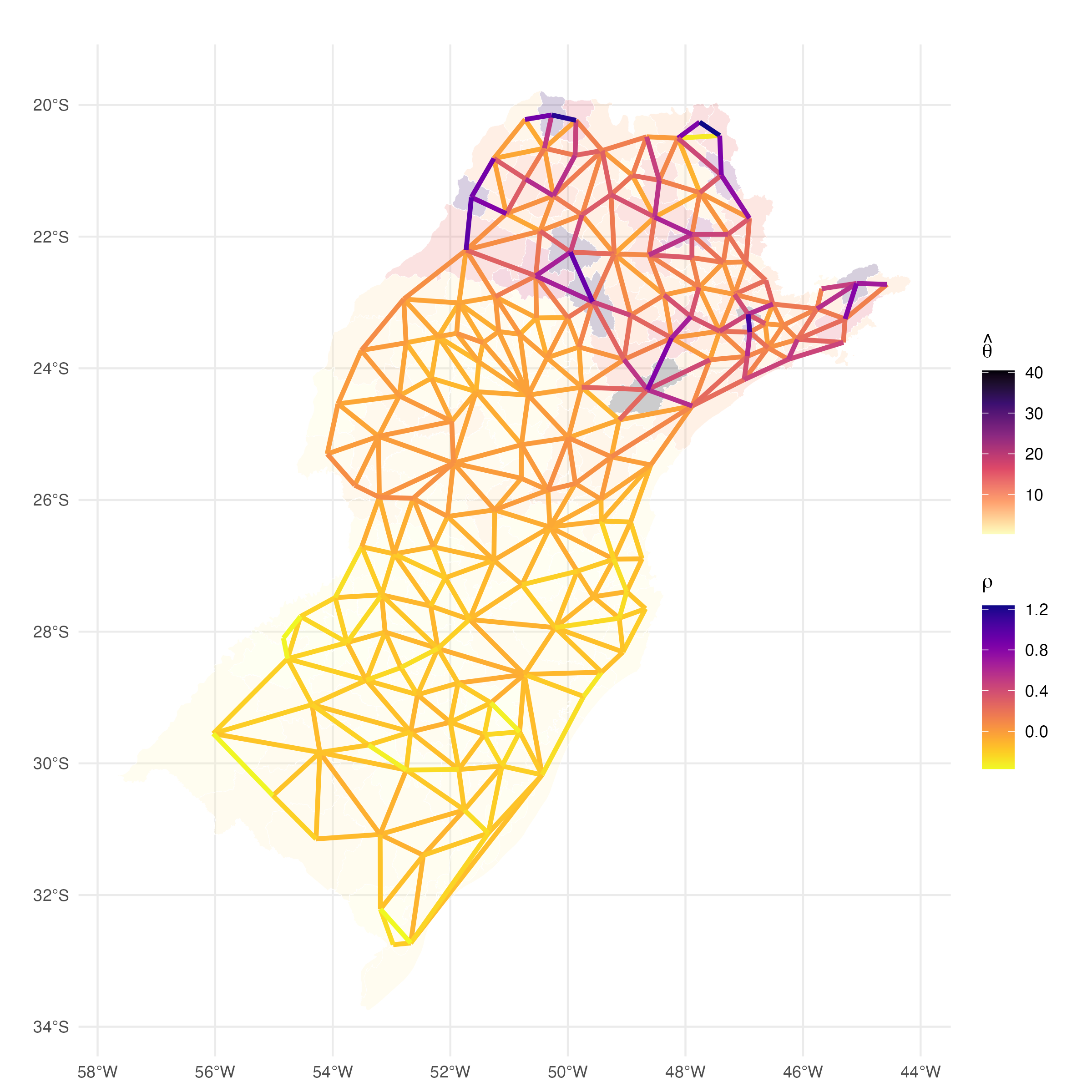}
\caption{Gaussian \modelname{} prior.}
\label{fig:simu_rho_gauss}
\end{subfigure}
\hfill
\begin{subfigure}{0.45\textwidth}
\centering
\includegraphics[width=\linewidth]{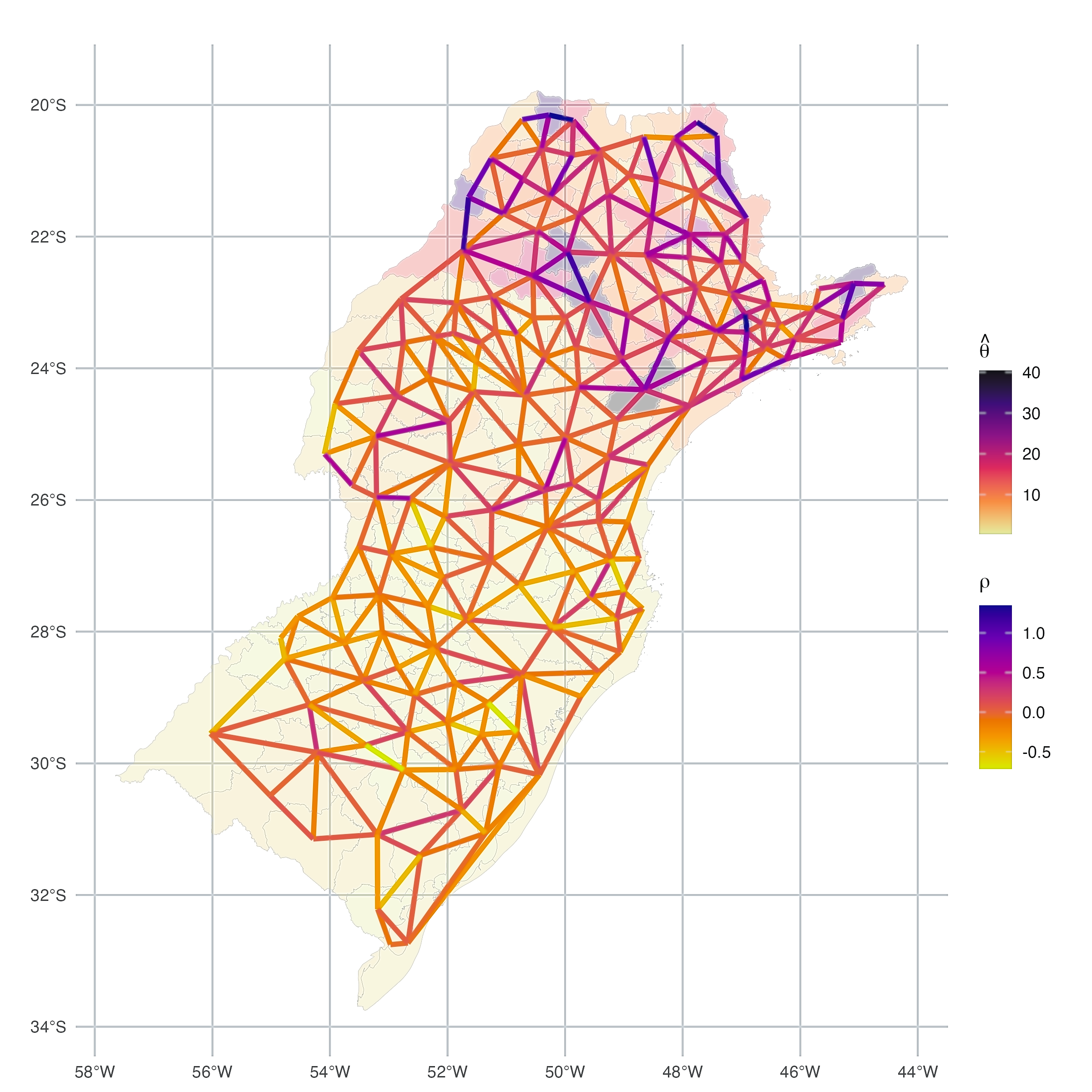}
\caption{\modelname{}--Skew prior.}
\label{fig:simu_rho_skew}
\end{subfigure}
\caption{Posterior medians of the edge-based latent effect $\rho$ and the spatial effect $\theta$ under the Gaussian \modelname{}  (a) and the \modelname{}--Skew  (b) prior distributions.}
\label{fig:simu_rho_both}
\end{figure}

Figure~\ref{fig:simu_rho_both} further clarifies the distinction between Gaussian and skew--normal edge-based priors. Under the Gaussian \modelname{} specification, the posterior median of the edge effects $\boldsymbol{\rho}$ varies smoothly across the domain, with gradual changes spread over many edges. Under \modelnamesn{}, the posterior median concentrates variation on the specific edges where skewness was introduced, yielding a localized and directionally consistent transition. This contrast highlights the role of the skew--normal perturbation as a mechanism for assigning asymmetric influence to selected adjacencies.

These qualitative differences are reflected in the quantitative model comparison reported in Table~\ref{tab:criterios_modelos}. The CAR model performs worse according to all criteria, reflecting its inability to represent the simulated directional structure efficiently. The Gaussian \modelname{} prior improves substantially by relocating dependence to the edge graph. The \modelnamesn{} model achieves the best overall performance, with lower information criteria and a smaller effective dimension, indicating that directional asymmetry is an essential feature of the latent field and can be captured parsimoniously through edge-based skewness.

\begin{table}[ht]
\centering
\caption{Comparison of model performance criteria for the CAR, \modelname{}, and \modelname{}--Skew models}
\label{tab:criterios_modelos}
\begin{tabular}{lllllll}
  \hline
Model & Dbar & pD & DIC & WAIC & LOOIC & RMSE \\ 
  \hline
CAR & 614.29 & 125.21 & 739.50 & 687.41 & 715.76 & 2.57 \\ 
\modelname{} & \textbf{580.82} & 93.86 & 674.68 & 642.70 & 671.94 & 2.45 \\ 
\modelname{}--Skew & 583.51 & \textbf{50.87} & \textbf{634.38} & \textbf{631.30} & \textbf{644.32} & \textbf{2.42} \\ 
  \hline
\end{tabular}
\end{table}

Taken together, these results clarify when a spatial prior may reasonably be regarded as skew--normal. Asymmetry must arise from a joint construction with an identifiable direction, propagate coherently through the graph, and admit a decomposition separating symmetric smoothing from directional perturbation. The \modelnamesn{} model satisfies these criteria by introducing skewness at the level where spatial transitions occur—edges—while preserving the interpretability and computational advantages of Gaussian edge-based models.

\section{ Spatial modeling of cancer incidence in Southern Brazil}
\label{aplicacion}

We illustrate the practical performance of the proposed methodology by applying the CAR, Gaussian \modelname{}, and \modelnamesn{} models to municipal-level cancer incidence data from Southern Brazil. The analysis focuses on two outcomes—lung cancer and colon cancer—which exhibit distinct spatial patterns and therefore provide a useful testbed for assessing the benefits of edge-based dependence and directional asymmetry.

The data consist of observed counts aggregated at the municipality level, together with expected counts derived from population-at-risk information obtained from the Brazilian public health system (DATASUS; {Informacoes de Saude Morbidade e Fatores de Risco}). Spatial dependence is defined through the municipal adjacency graph, from which we construct the node--edge incidence matrix and the associated line graph required for the Gaussian \modelname{} and \modelnamesn{} specifications. All three models are fitted within the same Poisson log--linear framework, using identical covariates and prior settings for regression parameters, so that differences in performance can be attributed primarily to the structure of the spatial prior.

Table~\ref{tab:comparison_all} summarizes the model comparison criteria for lung and colon cancer, reporting DIC, WAIC,  and RMSE. For lung cancer, the skew--normal edge-based model (\modelnamesn{}) achieves the lowest values for all likelihood-based criteria, indicating a clear improvement in predictive performance when directional asymmetry is allowed. The Gaussian \modelname{} model yields competitive RMSE values but produces a smoother and more isotropic latent field, while the CAR model performs noticeably worse across all criteria.


\begin{table}[ht]
\centering
\caption{Model comparison criteria for lung and colon cancer under CAR, \modelname{}, and \modelname{}-sk.}
\label{tab:comparison_all}
\begin{tabular}{lllllll}
\hline
\textbf{Cancer} & \textbf{Model} & \textbf{Dbar} & \textbf{pD} & \textbf{DIC} & \textbf{WAIC} &\textbf{RMSE} \\
\hline
Lung & CAR & 1475.51 & 158.73 & 1634.24 & 1587.09 & 35.64 \\
     & \modelname{} & 1476.69 & 156.06 & 1632.75 & 1587.55 & \textbf{35.54} \\
     & \modelname{}-sk & \textbf{1479.39} & \textbf{103.13} & \textbf{1582.52} & \textbf{1558.68} & 36.34 \\
\hline
Colon & CAR & 1161.52 & 178.23 & 1339.74 & 1279.57 & 19.32 \\
      & \modelname{} & 1158.40 & \textbf{174.58} & \textbf{1332.99} & 1273.64 & 19.66 \\
      & \modelname{}-sk & \textbf{1158.17} & 183.10 & 1341.28 & \textbf{1271.04} & \textbf{19.28} \\
\hline
\end{tabular}
\end{table}

For colon cancer, the differences between models are more subtle. The Gaussian \modelname{} specification attains the lowest DIC, suggesting an efficient balance between fit and complexity under a largely symmetric latent structure. Nevertheless, \modelnamesn{} outperforms both competitors in WAIC and RMSE, indicating that even modest departures from symmetry can be relevant for prediction. Taken together, these results suggest that while not all spatial fields require asymmetric modeling, allowing for skewness does not degrade performance and can yield tangible gains when directional effects are present.

\begin{figure}[htb]
\centering
\begin{subfigure}{0.48\textwidth}
    \centering
    \includegraphics[width=\textwidth]{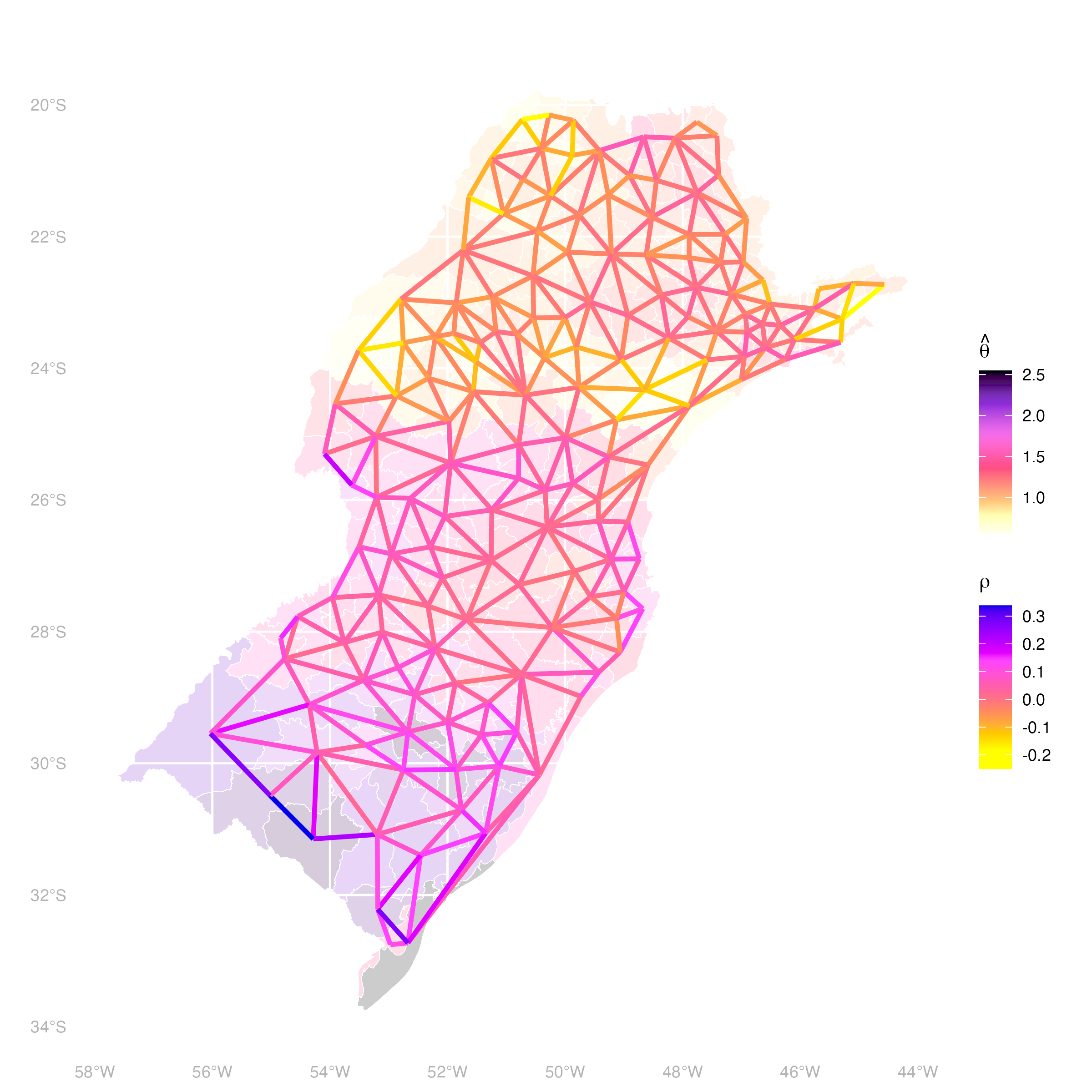}
    \caption{Gaussian \modelname{}: estimated spatial vector $\hat{\rho}$.}
    \label{fig:lung_rho_RENeGe}
\end{subfigure}
\hfill
\begin{subfigure}{0.48\textwidth}
    \centering
    \includegraphics[width=\textwidth]{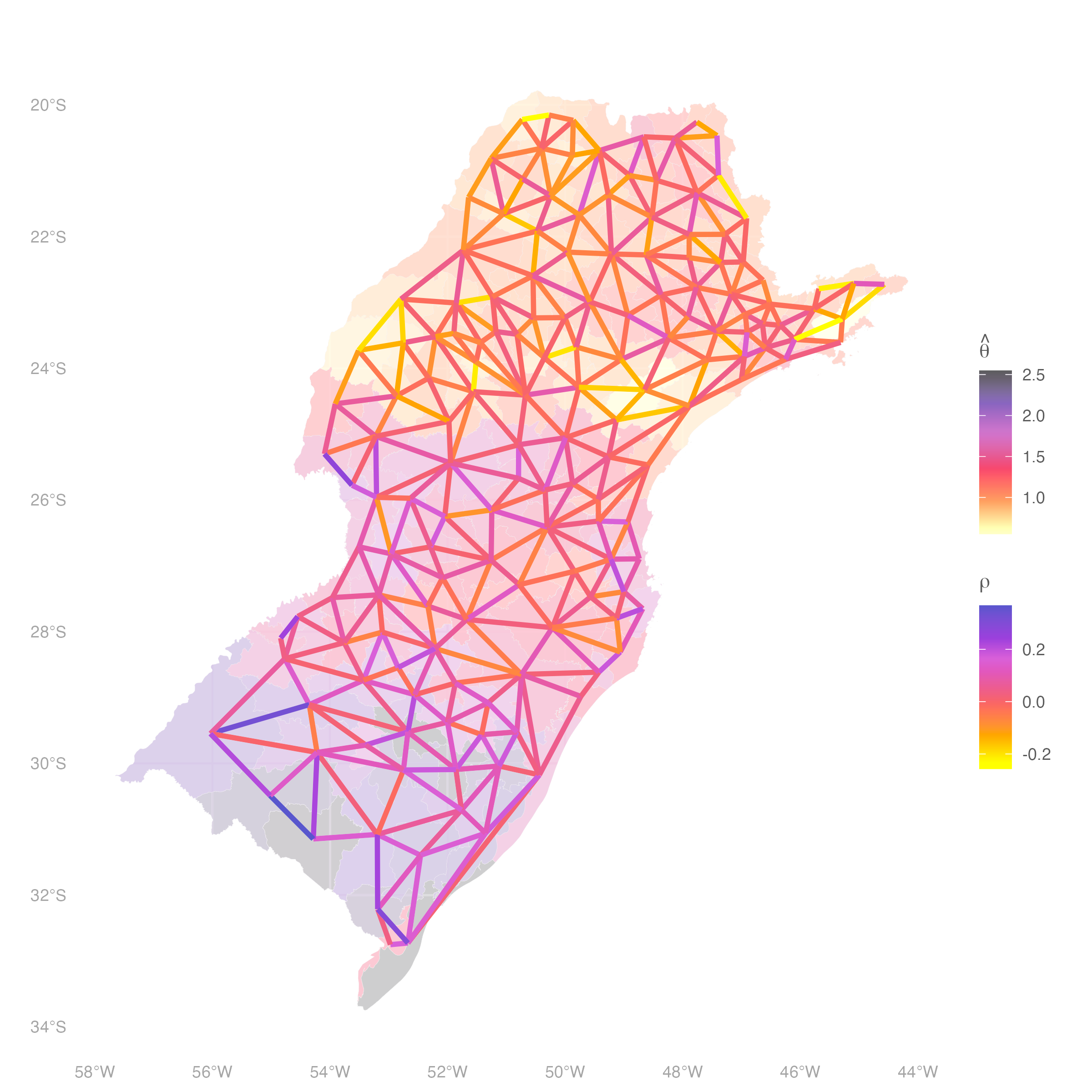}
    \caption{Skew-normal \modelnamesn{}: estimated spatial  $\hat{\rho}$.}
    \label{fig:lung_rho_sk}
\end{subfigure}
\caption{Comparison of the edges-based latent effects $\hat{\rho}$ (edges) and the spatial effect $\theta$  for lung cancer under the Gaussian (left) and skew-normal (right) \modelname{} models.}
\label{fig:lung_filters_comparison}

\centering
\begin{subfigure}{0.48\textwidth}
    \centering
    \includegraphics[width=\textwidth]{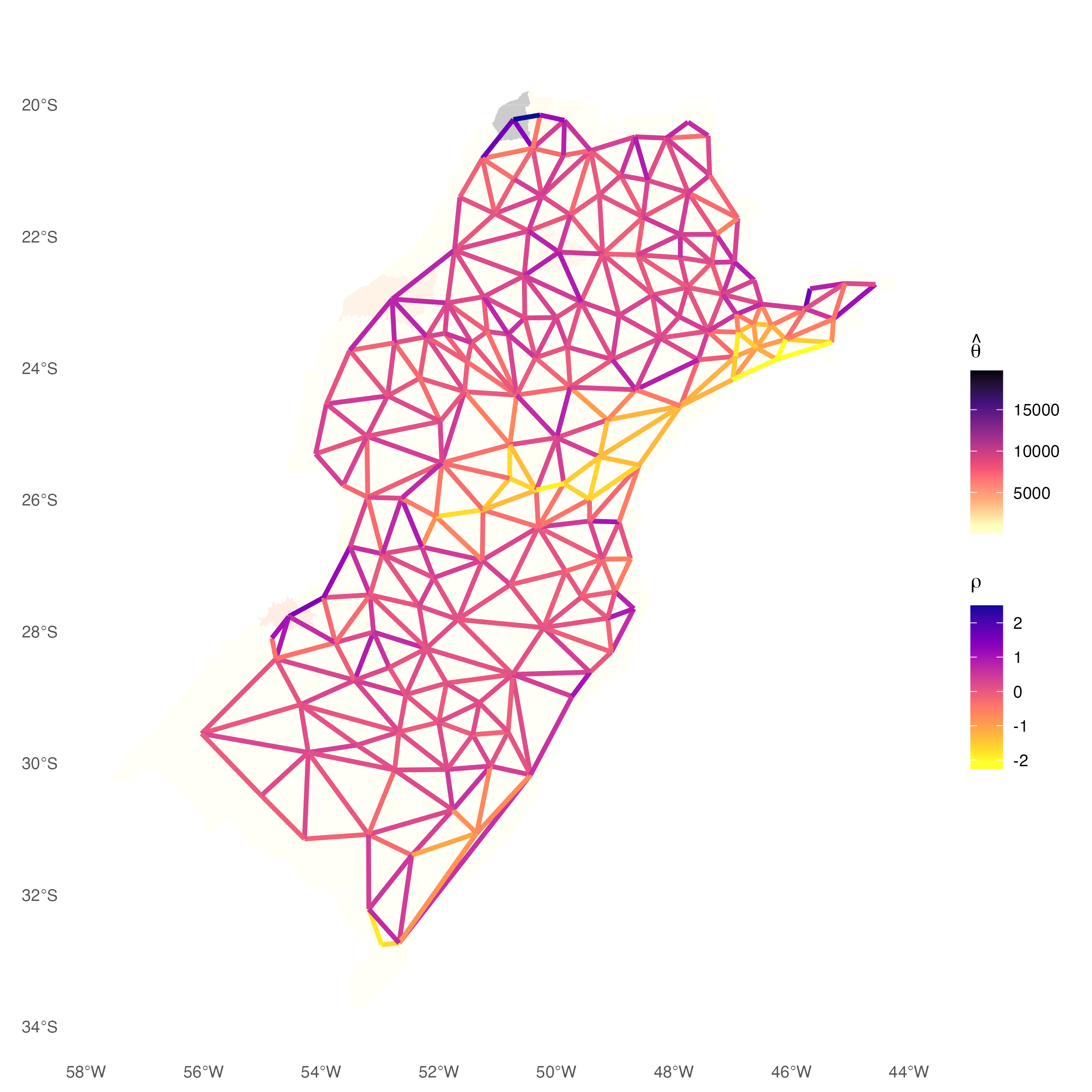}
    \caption{Gaussian \modelname{}: estimated spatial vector $\hat{\rho}$.}
    \label{fig:colon_rho_RENeGe}
\end{subfigure}
\hfill
\begin{subfigure}{0.48\textwidth}
    \centering
    \includegraphics[width=\textwidth]{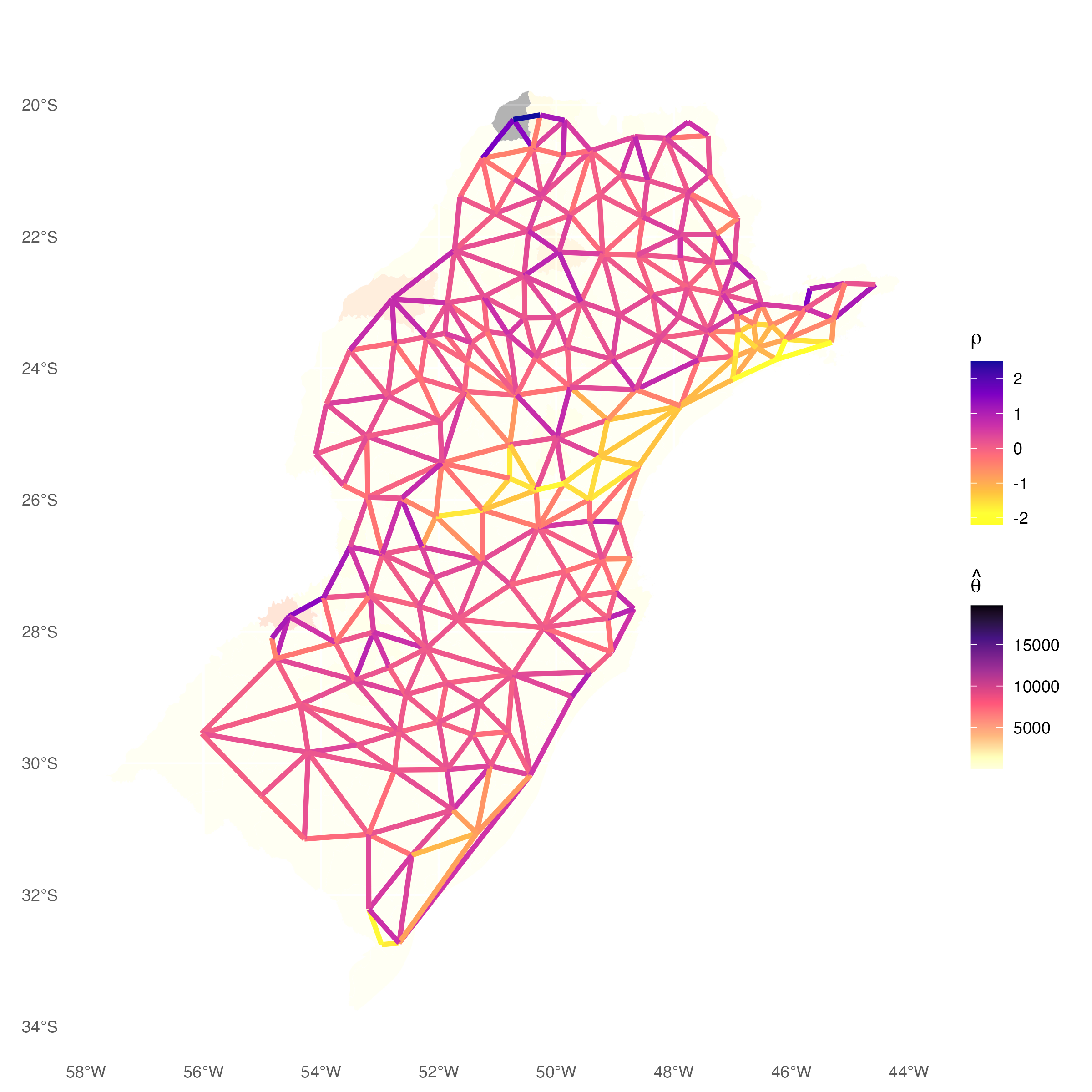}
    \caption{Skew-normal \modelnamesn{}: estimated spatial $\hat{\rho}$.}
    \label{fig:colon_rho_sk}
\end{subfigure}
\caption{Comparison of the edges-based latent effects $\hat{\rho}$ (edges) and the spatial effect $\theta$ for colon cancer under the Gaussian (left) and skew-normal (right) \modelname{} models.}
\label{fig:colon_filters_comparison}
\end{figure}

Additional insight is obtained by examining the posterior medians of the edge-level latent effects $\boldsymbol{\rho}$. Figures~\ref{fig:lung_rho_RENeGe} and \ref{fig:lung_rho_sk}  (resp., Figures~\ref{fig:colon_rho_RENeGe} and \ref{fig:colon_rho_sk}) display the estimated $\hat{\boldsymbol{\rho}}$ and $\hat{\boldsymbol{\theta}}$ surfaces for lung cancer (resp.,colon cancer) under the Gaussian \modelname{} and skew--normal \modelnamesn{} models, respectively. Because $\boldsymbol{\rho}$ is defined on the edges of the spatial graph, these maps provide a direct view of how dependence is transmitted across neighboring municipalities.

\clearpage

Under the Gaussian \modelname{} prior, the estimated edge effects vary smoothly over the graph, indicating symmetric propagation of spatial influence across adjacent regions. In contrast, the \modelnamesn{} specification yields sharper and more heterogeneous edge effects, with pronounced changes concentrated along specific subsets of edges. These localized patterns suggest asymmetric propagation of spatial structure—features that are systematically smoothed out under purely Gaussian formulations.

A similar contrast is observed for colon cancer (Figures~\ref{fig:colon_rho_RENeGe} and \ref{fig:colon_rho_sk}). Although the overall magnitude of spatial dependence is weaker for this outcome, the skew--normal model again reveals localized directional deviations in the edge effects that are absent from the Gaussian specification. This behavior reinforces the interpretation of \modelnamesn{} as a mechanism for capturing directional, edge-aligned transitions rather than merely increasing local variability.

Overall, the application demonstrates that modeling skewness directly on the edge graph provides a flexible and interpretable extension of Gaussian edge-based priors. When the underlying risk surface exhibits directional or boundary-aligned features, \modelnamesn{} yields improved predictive performance and sharper localization of spatial effects. At the same time, when asymmetry is weak or absent, the model remains competitive with symmetric alternatives, reducing the risk of overfitting. These properties make \modelnamesn{} a practical tool for spatial epidemiological analyses in which asymmetric or border-driven effects are substantively plausible.

\section{Conclusions}
\label{sec:conclusions}

We have introduced \modelnamesn{}, a skew--normal extension of the edge--based \modelname{} prior, designed to capture directional structure in spatial fields that symmetric Gaussian models cannot represent. The proposed construction combines a Gaussian dependence structure on the line graph with a parsimonious skewness perturbation defined on edges, yielding a spatial prior that remains interpretable while allowing asymmetric behavior aligned with specific adjacencies. The hierarchical formulation of the skewness vector, together with the auxiliary half--normal variable, ensures identifiability and enables the strength and orientation of asymmetry to be inferred from the data.

Through controlled simulation experiments, we demonstrated that \modelnamesn{} effectively recovers directional gradients when the underlying signal is concentrated along a subset of edges. In such settings, the skew--normal component produces sharper and more localized transitions than Gaussian alternatives, which tend to regularize abrupt changes symmetrically across the graph. Comparisons with CAR and Gaussian \modelname{} priors highlight that, although these models capture global smoothness, they are intrinsically limited in their ability to represent directional or boundary--aligned effects.

The empirical application to cancer incidence data further illustrates the practical relevance of the proposed approach. By operating directly on the edge graph, \modelnamesn{} preserves the geographic interpretation of spatial transitions while producing stable node--level risk surfaces. Improvements in likelihood--based criteria indicate that modeling asymmetry is not merely a cosmetic refinement, but a substantive component of spatial structure in certain applications. At the same time, the use of low--rank parameterizations and auxiliary--variable representations ensures that the method remains computationally tractable for graphs with many edges.

Overall, \modelnamesn{} provides a coherent and flexible framework for spatial modeling in settings where directional effects, edge--driven transitions, or asymmetric propagation of spatial influence are scientifically plausible. Future work may consider alternative low--rank representations for the skewness component, extensions to multivariate or spatiotemporal outcomes, and integration with non--Gaussian likelihoods relevant to environmental and epidemiological applications.

\section*{Acknowledgments}

\subsection*{Author contributions}

All authors contributed to the development of the model, the theoretical analysis, and the writing of the manuscript. The first author implemented the computational methods, conducted the simulation studies, collected the real-data sets, and carried out the empirical analysis.

\subsection*{Financial disclosure}

 This work was partially funded by the Brazilian agencies: Coordena\c c\~ao de Aperfei\c coamento de Pessoal de Ensino Superior (CAPES), Conselho Nacional de Desenvolvimento Cient\'{i}fico e Tecnológico (CNPq) and Funda\c c\~ao de Amparo \`a Pesquisa do Estado de Minas Gerais FAPEMIG.

\subsection*{Conflict of interest}

The authors declare no potential conflict of interests.

\section*{Supporting information}

The R code used to implement the proposed RENeGe-SK model, reproduces the
simulation studies, and generate the results reported in this paper is
available in a public GitHub repository at \url{https://github.com/DannaCruz/RENeGe-sk}.

\bibliography{wileyNJD-APA}%

\end{document}